\documentclass[preprint,showpacs,preprintnumbers,amsmath,amssymb]{revtex4}

\usepackage{graphicx}
\usepackage{dcolumn}
\usepackage{bm}
\usepackage{subfigure}
\usepackage{theorem}

\newcommand{\qed}{\hbox{\rule[-2pt]{3pt}{6pt}}}

\def\R{\mathbb{R}}
\def\C{\mathbb{C}}
\def\W{\mathbb{W}}
\def\bE{\mathbb{E}}

\def\U{\mathbb{U}}

\def\x{{\bf x}}
\def\y{{\bf y}}
\def\z{{\bf z}}
\def\X{{\bf X}}
\def\Z{{\bf Z}}
\def\T{{\bf T}}
\def\B{{\bf B}}

\def\sF{{\cal F}}
\def\sL{{\cal L}}
\def\sG{{\cal G}}
\def\sH{{\cal H}}
\def\sS{{\cal S}}
\def\sN{{\cal N}}

\def\1{{\bf 1}}

\def\vmu{\vec{\mu}}
\def\vlambda{\vec{\lambda}}
\def\vtau{\vec{\tau}}
\def\vnu{\vec{\nu}}

\begin{document}

\preprint{}

\title{O'Connell's process as a vicious Brownian motion}

\author{Makoto Katori}
\email{katori@phys.chuo-u.ac.jp}
\affiliation{%
Department of Physics,
Faculty of Science and Engineering,
Chuo University, 
Kasuga, Bunkyo-ku, Tokyo 112-8551, Japan 
}%

\date{10 December 2011}

\begin{abstract}
Vicious Brownian motion is a diffusion scaling limit
of Fisher's vicious walk model,
which is a system of Brownian particles in one dimension 
such that if two of them meet they kill each other.
We consider the vicious Brownian motion conditioned
never to collide with each other, and call it
the noncolliding Brownian motion.
This conditional diffusion process is 
equivalent to the eigenvalue
process of a Hermitian-matrix-valued 
Brownian motion studied by Dyson.
Recently O'Connell introduced 
a generalization of the noncolliding Brownian motion 
by using the eigenfunctions (the Whittaker functions)
of the quantum Toda lattice in order to analyze
a directed polymer model in 1+1 dimensions.
We consider a system of one-dimensional Brownian 
motions with a long-ranged killing term
as a generalization of the vicious Brownian motion
and construct the O'Connell process as a conditional
process of the killing Brownian motions
to survive forever.
\end{abstract}


\maketitle

\section{Introduction}

Vicious walk model introduced by Fisher \cite{Fis84}
is the system of one-dimensional random walkers
such that, if neighboring walkers meet, they kill each other.
Though the model looks sinister, what we are interested in
is to evaluate the probability that 
for a finite time-interval any neighboring
pair of vicious walkers do not meet and thus
all walkers survive;
in other words, the probability that the peace is kept 
\cite{GOV98,KGV00,KGV03}.
If we take appropriate continuum limit
(the diffusion scaling limit), we obtain 
``vicious Brownian motion"
(vicious BM) \cite{KT02,CK03}.
Assume that the number of particles of vicious BM is 
$N \geq 2$ and write the positions as
$x_j, 1 \leq j \leq N$.
Then the configuration space of them conditioned
never to collide is
\begin{equation}
\W_N=\{ \x =(x_1, x_2, \dots, x_N) 
\in \R^N: x_1 < x_2 < \cdots < x_N \},
\label{eqn:Weyl1}
\end{equation}
which is called the Weyl chamber of type A$_{N-1}$
in the representation theory \cite{FH91}.
The boundaries of this region $\partial \W_N$
in the $N$-dimensional real space $\R^N$
consists of the hyperplanes $x_j=x_{j+1}, 1 \leq j \leq N-1$,
each of which corresponds to 
occurrence of collision of the
$j$-th and $(j+1)$-th particles
in the vicious BM.
If we regard $\x$ as a position vector
of the $N$-dimensional BM within $\W_N$
and $\partial \W_N$ as an absorbing
boundary such that when a particle hit 
the boundary it is immediately absorbed,
the system is identified with the
absorbing BM in $\W_N$.
The harmonic function,
$\Delta h_N(\x) \equiv \sum_{j=1}^{N}
\partial^2 h_N(\x)/\partial x_j^2=0,
\x \in \W_N$, satisfying the
Dirichlet boundary condition
$h_N(\x)=0, \x \in \partial \W_N$,
is uniquely determined up to a constant
factor as
\begin{equation}
h_N(\x)=\prod_{1 \leq j < k \leq N}
(x_k-x_j), \quad
\x \in \W_N,
\label{eqn:hN1}
\end{equation}
which is equal to the Vandermonde 
determinant
$\det_{1 \leq j, k \leq N}
[x_j^{k-1}]$.
Then we can show that the survival probability
up to time $t \geq 0$ of $N$-particle system of vicious BM
starting from an initial configuration
$\x \in \W_N$ has the asymptotics
\begin{equation}
h_N \left( \frac{\x}{\sqrt{t}} \right)
= t^{-N(N-1)/4} h_N(\x)
\quad \mbox{in} \quad
\frac{|\x|}{\sqrt{t}} \to 0,
\label{eqn:asym1}
\end{equation}
where $|\x|=\sqrt{x_1^2+\cdots + x_N^2}$
(see, for instance, \cite{KT07}).

As a matter of course, the survival probability
decreases in time for any $\x \in \W_N$. 
The result (\ref{eqn:asym1})
implies, however, the decay is slow in the sense
that it is not exponential but follows
the power-law in time. This observation has led us
to study the system of BMs conditioned
never to collide with each other, 
which we call the {\it noncolliding BM}.
The fundamental properties of this process
are the following 
\cite{KT07}.
(i) Let the transition probability density
of a single BM be 
$p(t, y|x)=e^{-(y-x)^2/2t}/\sqrt{2 \pi t}$.
Then the transition probability density of the
absorbing BM in $\W_N$ from a configuration
$\x \in \W_N$ to $\y \in \W_N$ in time interval 
$t \geq 0$ is given by the Karlin-McGregor
determinant \cite{KM59}
\begin{equation}
q_N(t, \y|\x)=\det_{1 \leq j, k \leq N}
[ p(t, y_j|x_k)],
\label{eqn:KM1}
\end{equation}
or equivalently \cite{KT04} given by the
Harish-Chandra-Izykson-Zuber integral \cite{HC57,IZ80}
\begin{equation}
q_N(t,\y|\x) = 
\frac{t^{-N^2/2}}{c_N} h_N(\x) h_N(\y)
\int_{\U(N)}
e^{ - {\rm Tr}(\Lambda_{\tiny \x}-U^{\dagger} \Lambda_{\tiny \y} U)^2
/2 t} dU,
\label{eqn:HCIZ1}
\end{equation}
where  $dU$ is the Haar measure of the space of
unitary matrices $\U(N)$ normalized as
$\int_{\U(N)}dU=1$, 
$\Lambda_{\x}={\rm diag}(x_{1}, x_{2}, \dots, x_{N})$, 
$\Lambda_{\y}={\rm diag}(y_{1}, y_{2}, \dots, y_{N})$,
and $c_N=(2 \pi)^{N/2} \prod_{j=1}^{N} \Gamma(j)$
with the gamma function $\Gamma(j)$.
(ii) The transition probability density of the
noncolliding BM, $p_N(t, \y|\x)$, is then
given by the harmonic transform of $q_N(t, \y|\x)$
with (\ref{eqn:hN1}) in the sense of Doob \cite{Doo84},
\begin{equation}
p_N(t,\y|\x)=\frac{h_N(\y)}{h_N(\x)} 
q_N(t, \y|\x),
\quad \x, \y \in \W_N, \quad t \geq 0.
\label{eqn:h-trans1}
\end{equation}
(iii) If we regard (\ref{eqn:h-trans1}) as a function
of $t$ and initial configuration $\x$,
it is a solution of the following backward
Kolmogorov equation
\begin{eqnarray}
\frac{\partial}{\partial t} u(t,\x)
&=& \frac{1}{2} \Delta u(t,\x)
+ \nabla \log h_N(\x) \cdot
\nabla u(t, \x)
\nonumber\\
&=& \frac{1}{2} \sum_{j=1}^{N} 
\frac{\partial^2}{\partial x_j^2} u(t, \x)
+ \sum_{1 \leq j \leq N}
\sum_{1 \leq k \leq N : k \not= j}
\frac{1}{x_j-x_k} 
\frac{\partial}{\partial x_j} u(t, \x)
\label{eqn:Kolm1}
\end{eqnarray}
under the initial condition
$u(0, \x)=\delta(\x-\y)
\equiv \prod_{j=1}^{N} \delta(x_j-y_j)$.

The direct consequence of the above facts
is the following (see, for example,
\cite{KS91}).
Let $\X(t)=(X_1(t), \dots, X_N(t))$ be the $N$-particle
system of the noncolliding BM.
Then it solves the system of stochastic differential
equations (SDEs)
\begin{equation}
dX_j(t)=d B_j(t)
+\sum_{1 \leq k \leq N : k \not= j}
\frac{dt}{X_j(t)-X_k(t)},
\quad 1 \leq j \leq N, \quad t \geq 0,
\label{eqn:Dyson1}
\end{equation}
where $B_j(t), 1 \leq j \leq N, t \geq 0$,
are independent one-dimensional standard BMs.
Eq.(\ref{eqn:Dyson1}) is nothing but
the system of SDEs for Dyson's BM model
with the parameter $\beta=2$ \cite{Dys62}
studied in the random matrix theory \cite{Meh04,For10},
which we will simply call {\it the Dyson model}
in this paper.
That is, the noncolliding BM is equivalent to
the eigenvalue process of the Hermitian-matrix-valued
BM \cite{Gra99,KT04}.

Recently O'Connell introduced a family of
diffusion processes of $N$ particles
in one dimension,
$\Z^{\vmu}(t)=(Z^{\vmu}_1(t), \dots, Z^{\vmu}_N(t)),
t \geq 0$ with a parameter given by an
$N$-dimensional real vector 
$\vmu=(\mu_1, \mu_2, \dots, \mu_N) \in \R^N$,
in which the $j$-th particle has 
a constant drift $\mu_j$,
$1 \leq j \leq N$ \cite{OCo09}.
It is an extension of the Dyson model
(\ref{eqn:Dyson1}) in the sense that,
if we consider the scaled process
$\varepsilon \Z^{0}(t/\varepsilon^2), t \geq 0$ with $\varepsilon > 0$
for the case $\vmu=0$
and take the limit $\varepsilon \to 0$,
then the limit process is equivalent 
to $\X(t), t \geq 0$.
He showed that the process $\Z^{\vmu}(t), t \geq 0$
is associated with the quantum Toda lattice
with the Hamiltonian \cite{KL99,KL01,GKLO06}
\begin{equation}
\sH_N=-\frac{1}{2} \sum_{j=1}^{N}
\frac{\partial^2}{\partial x_j^2}
+\sum_{j=1}^{N-1} e^{-(x_{j+1}-x_j)}.
\label{eqn:qToda1}
\end{equation}
The O'Connell process is very rich in mathematics
connecting with quantum integrable systems,
representation theory of Lie groups/algebras,
the Whittaker functions, theory of intertwining
relations of Markov processes, and so on.
He discussed the importance of his process
to study a model of 1+1 dimensional directed
polymers in random environment with finite
temperature \cite{OCo09}.

The purpose of the present paper is to discuss
the O'Connell process as a generalized version
of vicious BM with appropriate conditions
at least for the special case $\vmu=0$.
(See 
\cite{GG10a,GG10b,BJ02,OY02,Joh04,War07,
SMCR08,KIK08,FN08,BFPSW09,KMW09,NM09,
FMS11,IK11,SK11,RS11} 
for other generalizations 
and recent topics of vicious BM
and noncolliding BM.
We note that interesting connections
between random growth models and
the Toda lattice Hamiltonian
is discussed in \cite{GNSS11}.)

The paper is organized as follows.
In Sec.II through the Feynman-Kac formula,
we introduce a system of Brownian particles
with the killing term which is in the same form
as the potential term in the quantum Toda lattice
Hamiltonian (\ref{eqn:qToda1}) 
and discuss it as a generalization
of the vicious BM.
In Sec.III the transition probability density
$Q_N(t, \y|\x)$ of the $N$-particle system of killing BMs
is expressed as an integral of a product
of eigenfunctions of the quantum Toda lattice
over the Sklyanin measure.
Then asymptotics of $Q_N(t,\y|\x)$ in $t \to \infty$
is estimated (Lemma 1).
In Sec.IV we introduce a drift $\vmu$ in our
$N$-particle system of killing BMs and 
define the transition probability density
of the killing BMs conditioned to survive
up to time $0< T < \infty$.
By taking the double limits
$T \to \infty$ and $\vmu \to 0$,
we obtain the transition probability density
$P_N(t, \y|\x)$ for the killing BMs with
$\vmu=0$ conditioned to survive forever.
The main theorem is given there (Theorem 2),
by which the equivalence between the present
conditional process and the O'Connell process
with $\vmu=0$ is concluded.
We discuss a one-dimensional diffusion
process studied by Matsumoto and Yor \cite{MY00,MY05}
in Sec.V
as a motion of relative coordinate
in the $N=2$ case of our process.
The Matsumoto-Yor process with $\mu=0$ is realized
as a one-dimensional killing BM
conditioned to survive forever.
In Sec.VI we discuss some distributions
obtained by setting the special initial
conditions.
Section VII is devoted to summary and concluding 
remarks.
Appendix \ref{chap:appendix_MY} 
is given for proving an asymptotics
used in Sec.V. Some details of the $N=2$ case
of the O'Connell process are given in
Appendix \ref{chap:N=2case}.

\section{Quantum Toda lattice and Feynman-Kac formula}

Let $N \in \{2,3, \dots\}$.
Consider the eigenvalue problem of the 
quantum Toda lattice Hamiltonian (\ref{eqn:qToda1}),
\begin{equation}
\sH_N \Psi_{\gamma}(\x)
=\gamma \Psi_{\gamma}(\x), \quad
\x \in \R^N.
\label{eqn:qToda2}
\end{equation}
For $\vlambda=(\lambda_1, \dots, \lambda_N) \in \C^N$,
the eigenfunctions of (\ref{eqn:qToda2}) with
eigenvalues
\begin{equation}
\gamma=- \frac{1}{2} \sum_{j=1}^{N} \lambda_j^2
\label{eqn:qToda3}
\end{equation}
have been extensively studied
\cite{KL99,KL01,GKLO06}, which are expressed by
$\psi_{\vlambda}^{(N)}(\x)$ in the present paper.
Let $\T$ denote a triangular array
with size $N$,
$\T=(T_{k,j}, 1 \leq j \leq k \leq N)$.
We consider that the $N(N-1)/2$ elements
$T_{k,j}$ of $\T$ are independent
variables and introduce a function of them as
\begin{eqnarray}
\sF^{(N)}_{\vlambda}(\T)
&=& \sum_{k=1}^{N} \lambda_{k}
\left( \sum_{j=1}^{k} T_{k,j}
-\sum_{j=1}^{k-1} T_{k-1, j} \right)
\nonumber\\
&& - \sum_{k=1}^{N-1} \sum_{j=1}^k
\Big\{ e^{-(T_{k,j}-T_{k+1,j})}
+e^{-(T_{k+1, j+1}-T_{k,j})} \Big\},
\label{eqn:qToda4}
\end{eqnarray}
which depends on $\vlambda=(\lambda_1, \dots, \lambda_N)$.
For a given $\x \in \R^N$, let $\Gamma_N(\x)$ be
the space of all real triangular arrays $\T$
with size $N$ conditioned
\begin{equation}
T_{N,j}=x_j, \quad 1 \leq j \leq N.
\label{eqn:qToda5}
\end{equation}
We write the integral of a function $f$
of $\T$ over $\Gamma_N(\x)$ as
\begin{equation}
\int_{\Gamma_N(\x)} f(\T) d \T
\equiv \prod_{k=1}^{N} \prod_{j=1}^k
\int_{-\infty}^{\infty} d T_{k,j} \,
f(\T) \prod_{\ell=1}^{N} \delta(T_{N,\ell}-x_{\ell}).
\label{eqn:qToda6}
\end{equation}
Then the integral representation of 
$\psi^{(N)}_{\vlambda}(\x)$ is given by
\begin{equation}
\psi_{\vlambda}^{(N)}(\x)
=\int_{\Gamma_N(\x)} 
e^{\sF_{\vlambda}^{(N)}(\T)} d \T.
\label{eqn:qToda7}
\end{equation}
This multivariate function is a version of
Whittaker function (see \cite{OCo09}
and references therein). 

As a stochastic version of the Schr\"odinger equation
of the quantum Toda lattice
(obtained by performing the Wick rotation
in the Schr\"odinger equation),
we consider the following diffusion equation
\begin{equation}
\frac{\partial}{\partial t} u(t, \x)
=\sL_N u(t, \x)
\label{eqn:Markov1}
\end{equation}
with the infinitesimal generator of the process
\begin{equation}
\sL_N \equiv - \sH_N
= \frac{1}{2} \Delta -V_N(\x),
\label{eqn:Markov2}
\end{equation}
where
\begin{equation}
V_N(\x)=\sum_{j=1}^{N-1} e^{-(x_{j+1}-x_j)}.
\label{eqn:Markov3}
\end{equation}
If we follow the method of separation of
variables by setting
$u(t,\x)=T(t) \Psi_{\gamma}(\x)$,
(\ref{eqn:Markov1}) is decomposed into
the equations
$$
\frac{dT(t)}{dt}=- \gamma T(t)
$$
and (\ref{eqn:qToda2}). Then we can conclude that
for any $\vlambda \in \C^N$,
\begin{equation}
\exp \left( \frac{t}{2} \sum_{j=1}^{N} \lambda_j^2 \right)
\psi_{\vlambda}^{(N)}(\x)
\label{eqn:Markov4}
\end{equation}
solves the diffusion equation (\ref{eqn:Markov1}).

In the context of quantum mechanics, the function
$V_N(\x)$ given by (\ref{eqn:Markov3}) plays,
as a matter of course, a role of
potential energy.
Then the quantum system prefers the state
$x_{j+1} > x_j$ to the state
$x_{j+1} < x_j, 1 \leq j \leq N-1$,
since the former has lower energy than the latter.
On the other hand, in the context of 
stochastic calculus, $-V_N(\x)$ term 
in the infinitesimal generator of the process (\ref{eqn:Markov2})
acts as a {\it killing term}.
We consider $N$ independent one-dimensional
standard BMs starting from 0,
$B_j(t), 1 \leq j \leq N$, and for
$\x \in \R^N$ set
$\B^{\x}(t)=\x+\B(t)$,
where each element $B_j^{x_j}(t) = x_j+B_j(t)$ is 
a one-dimensional standard BM starting from $x_j,
1 \leq j \leq N$.
Then the Feynman-Kac formula 
(see, for instance, \cite{KS91}) implies that
the function
\begin{equation}
Q_N(t, \y|\x)
= \bE \left[ \1(\B^{\y}(t)=\x)
\exp \left\{ - \int_0^{t} 
V_N(\B^{\y}(s)) ds \right\} \right]
\label{eqn:FK2}
\end{equation}
solves the diffusion equation 
(\ref{eqn:Markov1})
under the initial condition
\begin{equation}
Q_N(0, \y|\x)=\delta(\x-\y),
\label{eqn:FK3}
\end{equation}
where $\bE[\, \cdot \,]$ denotes
the expectation over all realizations
of $N$-dimensional Brownian paths,
$\{\B^{\y}(s): 0 \leq s \leq t \}$, 
starting from $\y$,
and $\1(\omega)$ is the indicator function
of the event $\omega$;
$\1(\omega)=1$ if $\omega$ is satisfied,
$\1(\omega)=0$ otherwise.
The function $Q_N(t,\y|\x)$ is the transition 
probability density of the process (\ref{eqn:Markov1})
from a configuration $\x$ to a configuration $\y$
in time interval $t \geq 0$.
In the Feynman-Kac formula (\ref{eqn:FK2}),
we consider a collection of all paths
of BM in $\R^N$ starting from $\y$ to $\x$.
(Though the time direction is backward,
it is irrelevant in calculation,
since BM is time-reversible.)
The point of this formula is the following.
In order to give the transition probability density
$Q_N(t, \y|\x)$, we have to put a weight
\begin{eqnarray}
w_N &=& \exp \left\{ - \int_0^t 
V_N(\B^{\y}(s)) ds \right\}
\nonumber\\
&=& \exp \left\{ - \sum_{j=1}^{N-1}
\int_0^t
e^{-(B_{j+1}^{y_{j+1}}(s)
-B_j^{y_j}(s))} ds \right\}
\label{eqn:FK4}
\end{eqnarray}
to each realization of path of the
$N$-dimensional BM and take 
a summation over all realizations of paths.
It is obvious that $w_N$ takes a real value in $[0, 1]$.
Then this summation of weighted paths
(a path integral) can be identified with 
a statistical-ensemble average of
Brownian paths, in which each path is
included in the ensemble with probability $w_N$ and is
deleted with probability $1-w_N$.
Deletion of an $N$-dimensional Brownian path
is interpreted as an event that the $N$-dimensional
BM is killed in the time interval $[0, t]$.
The weight $w_N$ is then regarded as 
the probability that the particle in $\R^N$ survives up to
time $t$.
(See Corollary 4.5 and explanation given below it
in Chapter 4 of \cite{KS91} for the
equivalence of the Feynman-Kac formula
to Brownian motion with killing of particles.)
Eq.(\ref{eqn:FK4}) gives the 
dependence of the survival probability 
on the realization of path 
$\{\B^{\y}(s), 0 \leq s \leq t\}$.
If the $N$-tuples of Brownian paths are ``well-ordered"
in the spatio-temporal plane,
$B_1^{y_1}(s) < B_2^{y_2}(s) < \cdots 
< B_N^{y_N}(s), 0 \leq s \leq t$,
and moreover
$B_{j+1}^{y_{j+1}}(s) \gg B_j^{y_j}(s),
1 \leq j \leq N-1, 0 \leq s \leq t$,
$w_N$ is large, 
while for a particle on the path 
$\{\B^{\y}(s), 0 \leq s \leq t\}$ in which
$B_{j+1}^{y_{j+1}}(s) < B_j^{y_j}(s), 1 \leq j \leq N$
for some $s \in[0, t]$, $w_N$ is small.

If we introduce a parameter $\varepsilon >0$,
then we can see that
\begin{eqnarray}
&& \lim_{\varepsilon \to 0} 
\exp \left\{ - \sum_{j=1}^{N-1}
\int_0^t
e^{-(B_{j+1}^{y_{j+1}}(s)
-B_j^{y_j}(s))/\varepsilon} ds \right\}
\nonumber\\
&& \qquad
= \1 \Big(
\mbox{
$B_1^{y_1}(s), \dots, B_N^{y_N}(s)$ 
do not collide
during $[0, t]$ } \Big)
\nonumber\\
&& \qquad 
=\1 \Big(
\B^{\y}(s) \in \W_N, 0 \leq \forall s \leq t \Big).
\label{eqn:FK5}
\end{eqnarray}
In this sense, the process (\ref{eqn:Markov1})
with (\ref{eqn:Markov2}) and (\ref{eqn:Markov3})
is an $N$-particle
system of killing BMs, which can be regarded as
an extension of the absorbing BM in $\W_N$.
In the next section, we explain
how to express the transition probability
density given by the
Feynman-Kac formula (\ref{eqn:FK2}) 
as a superposition of the Toda lattice
eigenfunctions (\ref{eqn:Markov4}).

\vskip 0.5cm
\noindent{\bf Remark.} \,
If we consider the present process not as 
an $N$-dimensional BM in $\R^N$ but 
as an $N$-particle system of one-dimensional BMs,
(\ref{eqn:FK2}) gives the transition probability density
in the case that mutual killing of particles
does not occur at all in time duration $t$,
since $\x$ and $\y$ are both $N$-particle configurations,
$\x, \y \in \R^N$.
In order to discuss processes, in which 
mutual killing of particles actually occurs
and total number of particles decreases in time,
we have to specify the way how to choose
pair of particles which are annihilated; 
{\it e.g.} the pair $(j,j+1)$
attaining $\min \{B_{k+1}^{y_{k+1}}(t)-B_{k}^{y_k}(t)\}$
is chosen. Note that in the original vicious BM,
colliding pairs of particles are pair annihilated.
In the present paper, however, we are interested in
the process conditioned that all $N$ particles
survive.

\section{Transition probability density and
its long-term asymptotics}

The problem which is discussed here is 
how to determine the function $g_{\vlambda}(\y)$
of $\vlambda \in \C^N, \y \in R^N$ and 
a subset $\Sigma$ of $\C^N$ such that
the integral of (\ref{eqn:Markov4}) 
\begin{equation}
\int_{\Sigma} \exp \left( \frac{t}{2}
\sum_{j=1}^{N} \lambda_j^2 \right)
\psi_{\vlambda}^{(N)}(\x) g_{\vlambda}(\y) d \vlambda
\label{eqn:superposition}
\end{equation}
is equal to $Q_N(t, \y|\x)$ given by
(\ref{eqn:FK2}).
This problem is solved by applying 
the theory of the Sklyanin measure \cite{Skl85}
defined by,
\begin{equation}
s_N(\vlambda) d \vlambda
\equiv \frac{1}{(2 \pi i)^N N!}
\prod_{1 \leq j < k \leq N}
\left\{ (\lambda_k-\lambda_j)
\frac{\sin \pi (\lambda_j-\lambda_k)}{\pi} \right\}
\prod_{\ell=1}^N d \lambda_{\ell}
\label{eqn:Skl1}
\end{equation}
for $\vlambda \in \Sigma =(i \R)^N$,
where $i=\sqrt{-1}$.
As a multi-dimensional extension of the fact that
Macdonald's functions of imaginary order
$\lambda \in i \R$, $K_{\lambda}(e^{-x})$, 
make complete basis of a suitable
set of functions with respect to the
measure $s_1(\lambda) d \lambda
=(i/\pi)^2 \lambda \sin(\pi \lambda)
d \lambda$
(see p.131 of \cite{Leb65} and Sec.V
in the present paper);
\begin{equation}
\frac{i}{\pi^2} \int_{-i \infty}^{i \infty} 
K_{\lambda}(e^{-x}) K_{\lambda}(e^{-y})
\lambda \sin (\pi \lambda) d \lambda
=\delta(x-y),
\label{eqn:Leb1}
\end{equation}
$x, y \in \R$, the following is valid
\cite{KL99,KL01},
\begin{equation}
\int_{(i\R)^N}
\psi_{\vlambda}^{(N)}(\x) \psi_{-\vlambda}^{(N)}(\y)
s_N(\vlambda) d \vlambda
=\delta(\x-\y),
\label{eqn:Lev2}
\end{equation}
$\x, \y \in \R^N$.
Note that if $\vlambda \in (i \R)^N$
then $\psi_{-\vlambda}^{(N)}(\y)$ is the
complex conjugate of $\psi_{\vlambda}^{(N)}(\y)$.

Since the transition probability density (\ref{eqn:FK2})
is a unique solution of the diffusion equation 
(\ref{eqn:Markov1}) satisfying the initial condition
(\ref{eqn:FK3}), the following is concluded;
\begin{equation}
Q_N(t, \y|\x)
=\int_{(i \R)^N} e^{t \sum_{j=1}^N \lambda_j^2/2}
\psi_{\vlambda}^{(N)}(\x)
\psi_{-\vlambda}^{(N)}(\y) s_N(\vlambda) d \vlambda.
\label{eqn:tpd1}
\end{equation}
It should be noted that Lemma 4.6 in 
\cite{KL99} implies
\begin{equation}
\int_{\R^{N}} \psi_{-\vtau}^{(N)}(\x)
\psi_{\vlambda}^{(N)}(\x) d \x
=\frac{1}{s_N(\vlambda)}
\frac{1}{N!} \sum_{\sigma \in \sS_N}
\delta(\sigma(\vlambda)-\vtau)
\label{eqn:ortho1}
\end{equation}
for $\vlambda, \vtau \in (i \R)^N$,
where $\sS_N$ is a set of all
permutations of $N$ indices
and $\sigma(\vlambda) \equiv
(\lambda_{\sigma(1)}, \dots, \lambda_{\sigma(N)})$
for $\sigma \in \sS_N$.
The orthogonality (\ref{eqn:ortho1}) guarantees the
Chapman-Kolmogorov equation
for the transition probability density
\begin{equation}
\int_{\R^N} Q_N(t, \z|\y) Q_N(s, \y|\x) d \y
=Q_N(t+s, \z|\x),
\quad \x, \z \in \R^N
\label{eqn:CK1}
\end{equation}
for $0 \leq s, t < \infty$.
As a matter of course, (\ref{eqn:FK2})
should satisfy it by the Markov property
of BMs.

If we change the integral variables in 
(\ref{eqn:tpd1}) with (\ref{eqn:Skl1}) as
$\lambda_j \mapsto \nu_j$ by
$\lambda_j \sqrt{t/2} = i \nu_j, 1 \leq j \leq N$,
we obtain the following expression for $Q_N(t, \y|\x)$,
\begin{eqnarray}
&& Q_N(t, \y|\x)
= \frac{2^{N^2/2}}{(2 \pi)^N N!} t^{-N^2/2}
\psi_0^{(N)}(\x) \psi_0^{(N)}(\y)
\nonumber\\
&& \times
\int_{\R^N} e^{-|\vnu|^2}
\prod_{1 \leq j < k \leq N}
\left[ (\nu_k-\nu_j) 
\frac{\sinh \{\pi \sqrt{2/t} (\nu_k-\nu_j)\}}
{\pi \sqrt{2/t}} \right]
\frac{\psi^{(N)}_{i \sqrt{2/t} \vnu}(\x)}
{\psi^{(N)}_0(\x)}
\frac{\psi^{(N)}_{-i \sqrt{2/t} \vnu}(\y)}
{\psi^{(N)}_0(\y)} 
d \vnu,
\label{eqn:QNexp1}
\end{eqnarray}
$\x, \y \in \R^N, t \geq 0$.
This integral expression should be compared with
the Schur function expansion of 
(\ref{eqn:KM1}) \cite{KT04},
\begin{eqnarray}
&& q_N(t, \y|\x) = 
\frac{t^{-N^2/2}}{(2 \pi)^{N/2}}
h_N(\x) h_N(\y)
\nonumber\\
&& \times \sum_{\nu: \ell(\nu) \leq N}
\prod_{j=1}^{N} \frac{1}{(\nu_j+N-j)!}
\exp \left( -\frac{|\x|^2+|\y|^2}{2t} \right)
s_{\nu}(\x/\sqrt{t}) s_{\nu}(\y/\sqrt{t}),
\label{eqn:qNexp1}
\end{eqnarray}
where 
$\{\nu=(\nu_1, \nu_2, \dots): \nu_1 \geq \nu_2 \geq \dots \}$ 
are partitions of integers,
$\ell(\nu)$ denotes the length of $\nu$,
and $s_{\nu}(\x)$ is the Schur function \cite{FH91}.

We can prove the following asymptotics
of the transition probability density.

\vskip 0.5cm
\noindent{\bf Lemma 1} \quad
Let
\begin{equation}
\alpha_N=\frac{N^2}{2}
\label{eqn:asymB1}
\end{equation}
and
$C_N=\prod_{n=1}^{N} \Gamma(n)/(2\pi)^{N/2}$.
Then
\begin{equation}
\lim_{t \to \infty}
\frac{t^{\alpha_N}}{C_N}
Q_N(t, \y|\x)
=\psi_{0}^{(N)}(\x) \psi_{0}^{(N)}(\y)
\label{eqn:asymB2}
\end{equation}
for $\x, \y \in \R^N$.
\vskip 0.5cm
\noindent{\it Proof.} \quad
In the limit $t \to \infty$, (\ref{eqn:QNexp1}) 
will behave as 
\begin{equation}
Q_N(t, \y|\x)
\simeq \frac{2^{N^2/2}}{(2 \pi)^N N!} t^{-N^2/2}
\psi^{(N)}_0(\x) \psi^{(N)}_0(\y)
\int_{\R^N} e^{-|\vnu|^2} 
\prod_{1 \leq j < k \leq N}
(\nu_k-\nu_j)^2 d \vnu.
\label{eqn:Qasym}
\end{equation}
The integral is 
a version of the Selberg integral
(a special case with $\gamma=1$ and $a=1$
in Eq. (17.6.7) of \cite{Meh04}),
\begin{equation}
\int_{\R^N}  e^{-|\vnu|^2}
\prod_{1 \leq j < k \leq N} (\nu_k-\nu_j)^2
=(2 \pi)^{N/2} 2^{-N^2/2} \prod_{n=1}^{N} n!.
\label{eqn:JN1}
\end{equation}
Then the lemma is obtained. \qed

\section{O'Connell process as an $N$-particle system of 
killing Brownian motions conditioned to survive forever}

Let $\vmu=(\mu_1, \dots, \mu_N) \in \R^N$
and introduce a drift term in the diffusion equation
\begin{equation}
\frac{\partial}{\partial t} u^{\vmu}(t,\x)
=\sL^{\vmu}_N u^{\vmu}(t, \x)
\label{eqn:drift1}
\end{equation}
with
\begin{eqnarray}
\sL_{N}^{\vmu} &=& \sL_N
-\vmu \cdot \nabla
\nonumber\\
&=& \sL_N
-\sum_{j=1}^{N} \mu_j \frac{\partial}{\partial x_j}.
\label{eqn:drift2}
\end{eqnarray}
It is easy to confirm that if $u(t,\x)$
solves (\ref{eqn:Markov1}), then
\begin{eqnarray}
u^{\vmu}(t, \x)
&=& \exp \left( - \frac{t}{2} |\vmu|^2 
+\vmu \cdot \x \right) u(t, \x)
\nonumber\\
&=& \exp \left( - \frac{t}{2} \sum_{j=1}^{N} \mu_j^2 
+\sum_{j=1}^{N} \mu_j x_j \right)
u(t, \x)
\label{eqn:drift3}
\end{eqnarray}
solves (\ref{eqn:drift1}).
The formula (\ref{eqn:drift3}) is called
the drift transformation from $u(t, \x)$
to $u^{\vmu}(t, \x)$ \cite{KS91}.
Then we can prove that the transition probability 
density for the process with drift $\vmu$ is given by
\begin{equation}
Q^{\vmu}_N(t, \y|\x)
= \exp \left\{ -\frac{t}{2}|\vmu|^2 
+ \vmu \cdot (\x-\y) \right\}
Q_N(t, \y|\x)
\label{eqn:drift4}
\end{equation}
for $\x, \y \in \R^N, t \geq 0, \vmu \in \R^N$.

\begin{figure}
\includegraphics[width=0.6\linewidth]{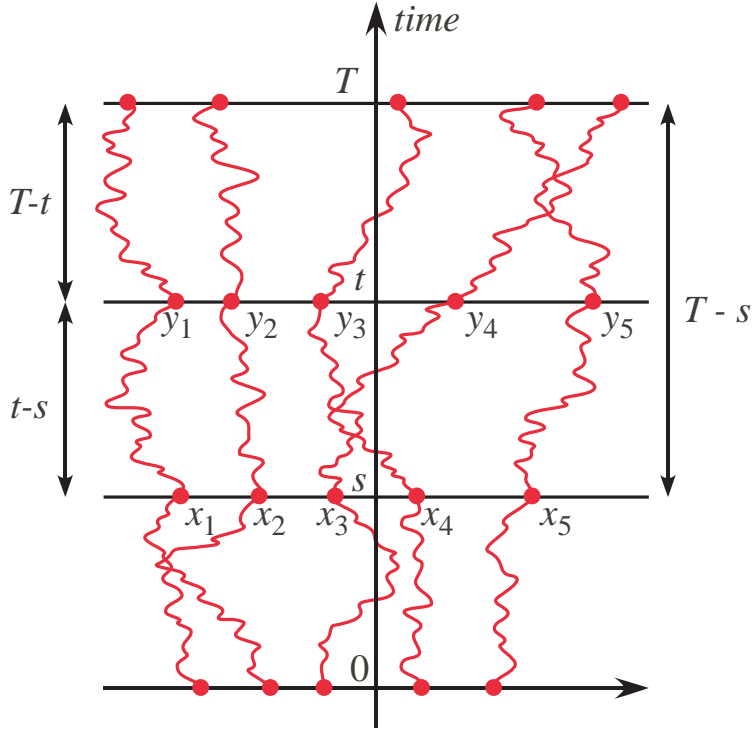}
\caption{An illustration of the paths in the case $N=5$ and $\vmu=0$
such that all particles survive up to time $T$, in which
particle configurations at times $s$ and $t$,
$0 \leq s \leq t \leq T$, are specified to be
$\x=(x_1, x_2, \dots, x_5)$ and 
$\y=(y_1, y_2, \dots, y_5)$, respectively.
By the Markov property of BMs, the behavior of paths
after time $s$ is independent of that before time $s$.
The total survival probability of the process
during time interval $[s,T]$ is
$\sN^{\vmu}_N(T-s, \x)$ under the condition that
the configuration at time $s$ is $\x$.
If we fix the configuration at time $t$
to be $\y$, $s \leq t \leq T$, 
the survival probability density is given by a product
$\sN^{\vmu}_N(T-t, \y) Q^{\vmu}_N(t-s, \y|\x)$,
since the events before time $t$ and after time $t$
are independent by the Markov property of BMs.
As shown by (\ref{eqn:PN1}), the ratio of them
gives the transition probability density
$P^{\vmu}_{N,T}(s, \x; t, \y)$.
}
\label{fig:Fig1}
\end{figure}

For $\x \in \R^N, 0 \leq T < \infty$,
let
\begin{equation}
\sN^{\vmu}_N(T, \x)
=\int_{\R^N} Q^{\vmu}_N(T, \y|\x) d \y,
\label{eqn:NN1}
\end{equation}
which gives the probability that
all $N$ particles survive up to time $T$ 
in the process of killing BMs with drift $\vmu$.

Now we consider the killing BM with drift $\vmu$
conditioned to survive up to time $T$.
Assume that $0 \leq s \leq t \leq T$
and let $P^{\vmu}_{N,T}(s, \x; t, \y)$ denote
the transition probability density 
from a configuration $\x$ at time $s$ to
a configuration $\y$ at time $t$
of this conditional process
as illustrated by Fig.\ref{fig:Fig1}.
By the Markov property of BMs,
we treat the paths only after time $s$
for $P^{\vmu}_{N,T}(s, \x; t, \y)$.
Since configurations at times
$s$ and $t$ are both specified to be
$\x$ and $\y$, but configuration
at the final time $T$ is arbitrary,
$P^{\vmu}_{N,T}(s, \x; t, \y)$
should be proportional to a product
of $Q^{\vmu}_N(t-s, \y|\x)$
and $\sN^{\vmu}_N(T-t, \y)$.
The product should be divided by
$\sN^{\vmu}_N(T-s, \x)$ to give
$P^{\vmu}_{N,T}(s, \x; t, \y)$, since
$\int_{\R^N} P^{\vmu}_{N,T}(s, \x; t, \y) d \y=1$
should hold for any $\x \in \R^N$.
Here we can see that
by the Chapman-Kolmogorov equation (\ref{eqn:CK1}),
$\int_{\R^N} \sN^{\vmu}_N(T-t, \y)
Q_N^{\vmu}(t-s, \y|\x) d \y
=\sN^{\vmu}_N(T-s, \x)$.
That is, we have the formula
\begin{equation}
P^{\vmu}_{N,T}(s, \x; t, \y)
=\frac{\sN^{\vmu}_N(T-t, \y)
Q^{\vmu}_{N}(t-s, \y|\x)}
{\sN^{\vmu}_N(T-s, \x)}.
\label{eqn:PN1}
\end{equation}
By definition, this conditional process
is a temporally inhomogeneous process,
which is clarified by the fact that
the transition probability density (\ref{eqn:PN1})
is the function not only of the time difference
$t-s$ but also of $T-s$ and $T-t$.

In order to obtain the temporally homogeneous
process, we take the limit $T \to \infty$.
Here we note that $\psi_0^{(N)}(\y)$ is the
eigenfunction of the quantum Toda lattice (\ref{eqn:qToda1}) 
with zero eigenvalue.
If $y_{j+1} \gg y_j$ for all $1 \leq j \leq N-1$,
the potential energy (\ref{eqn:Markov3}) becomes
zero and then $\psi_0^{(N)}(\y)$ behaves
similar to the harmonic function $h_N(\y)$.
On the other hand, it is known that
$\psi_{\vlambda}^{(N)}(\y) \sim \exp \{-2 e^{-(y_{j+1}-y_j)}\} 
\to 0$
as $y_{j+1}-y_j \to - \infty$, $1 \leq j \leq N-1$
for any $\vlambda \in (i \R)^N$ \cite{KL99}.
Then if $\vmu$ is chosen so that
$\vmu \in \W_N$, then
the integral $\int_{\R^N} \psi_0^{(N)}(\y) \exp(-\vmu \cdot \y) d\y$
is finite. In such a case,
by Lemma 1 and the drift transformation (\ref{eqn:drift4}), 
\begin{equation}
\lim_{T \to \infty}
\frac{T^{\alpha_N}}{C_N} e^{|\vmu|^2 T/2}
\sN^{\vmu}_N(T-s, \x)
= \exp(\vmu \cdot \x) \psi_{0}^{(N)}(\x) I^{\vmu}_0(N),
\label{eqn:NN2}
\end{equation}
$0 < \forall s < \infty$, 
with
\begin{equation}
I^{\vmu}_0(N)= \int_{\R^N} \psi_0^{(N)}(\y)
\exp (-\vmu \cdot \y) d \y,
\quad \vmu \in \W_N.
\label{eqn:NN2b}
\end{equation}
Then we have
\begin{eqnarray}
P^{\vmu}_N(t-s, \y|\x)
&\equiv& \lim_{T \to \infty}
P^{\vmu}_{N,T}(s, \x; t, \y)
\nonumber\\
&=& \exp \{-\vmu \cdot (\x-\y) \}
\frac{\psi_0^{(N)}(\y)}{\psi_0^{(N)}(\x)}
Q^{\vmu}_N(t-s, \y|\x)
\nonumber\\
&=& e^{-|\vmu|^2 t/2} 
\frac{\psi_0^{(N)}(\y)}{\psi_0^{(N)}(\x)}
Q_N(t-s, \y|\x)
\label{eqn:PN2}
\end{eqnarray}
for $0 \leq s \leq t < \infty$, $\x, \y \in \R^N, \vmu \in \W_N$.

Finally we take the limit $\vmu \to 0, \vmu \in \W_N$.
The obtained transition probability density
is given by
\begin{eqnarray}
P_N(t, \y|\x)
&\equiv& \lim_{\vmu \to 0, \vmu \in {\rm W}_N}
P^{\vmu}_N(t, \y|\x)
\nonumber\\
&=& \frac{\psi_0^{(N)}(\y)}{\psi_0^{(N)}(\x)}
Q_N(t, \y|\x),
\label{eqn:PN3}
\end{eqnarray}
$\x, \y \in \R^N, t \geq 0$.
It is an extension of (\ref{eqn:h-trans1}),
where the Vandermonde determinant
$h_N(\x)$ is replaced by the
eigenfunction $\psi^{(N)}_0(\x)$
with zero eigenvalue of the quantum Toda lattice.
Now we state the main theorem of the present paper.

\vskip 0.5cm
\noindent{\bf Theorem 2} \quad
The function (\ref{eqn:PN3}),
which is obtained as the transition probability
density of the killing BM conditioned
to survive forever, solves
the following differential equation,
\begin{eqnarray}
&& \frac{\partial}{\partial t} u(t, \x)
= \frac{1}{2} \Delta u(t, \x)
+\nabla \log \psi_0^{(N)}(\x) \cdot \nabla u(t, \x)
\nonumber\\
&& \qquad = \frac{1}{2} \sum_{j=1}^{N}
\frac{\partial^2}{\partial x_j^2} u(t, \x)
+\sum_{j=1}^N
\frac{\partial \log \psi_0^{(N)}(\x)}{\partial x_j}
\frac{\partial}{\partial x_j} u(t, \x)
\label{eqn:OConnell1}
\end{eqnarray}
under the initial condition
\begin{equation}
u(0, \x)=\delta(\x-\y).
\label{eqn:OConnell2}
\end{equation}
\vskip 0.5cm
\noindent{\it Proof.} \quad
By (\ref{eqn:PN3}), 
\begin{eqnarray}
&& \frac{\partial}{\partial x_j} 
P_N(t, \y|\x)
= -\frac{\psi_0^{(N)}(\y)}{(\psi_0^{(N)}(\x))^2}
\frac{\partial \psi_0^{(N)}(\x)}{\partial x_j}
Q_N(t, \y|\x)
+ \frac{\psi_0^{(N)}(\y)}{\psi_0^{(N)}(\x)}
\frac{\partial}{\partial x_j}
Q_N(t, \y|\x),
\nonumber\\
&& \frac{\partial^2}{\partial x_j^2} P_N(t, \y|\x)
= \frac{2 \psi_0^{(N)}(\y)}{(\psi_0^{(N)}(\x))^3}
\left( \frac{\partial \psi_0^{(N)}(\x)}{\partial x_j} \right)^2
Q_N(t,\y|\x)
\nonumber\\
&& \quad - \frac{\psi_0^{(N)}}{(\psi_0^{(N)}(\x))^2}
\frac{\partial^2 \psi_0^{(N)}(\x)}{\partial x_j^2} Q_N(t, \y|\x)
-2 \frac{\psi_0^{(N)}(\y)}{(\psi_0^{(N)}(\x))^2}
\frac{\partial \psi_0^{(N)}(\x)}{\partial x_j}
\frac{\partial}{\partial x_j} Q_N(t, \y|\x)
\nonumber\\
&& \quad + \frac{\psi_0^{(N)}(\y)}{\psi_0^{(N)}(\x)}
\frac{\partial^2}{\partial x_j^2} Q_N(t, \y|\x).
\label{eqn:Th1p1}
\end{eqnarray}
Since
$$
\frac{\partial \log \psi_0^{(N)}(\x)}{\partial x_j}
=\frac{1}{\psi_0^{(N)}(\x)} 
\frac{\partial \psi_0^{(N)}(\x)}{\partial x_j},
$$
RHS of (\ref{eqn:OConnell1}) is equal to
\begin{equation}
-\frac{\psi_0^{(N)}(\y)}{(\psi_0^{(N)}(\x))^2}
Q_N(t, \y|\x) \frac{1}{2} \sum_{j=1}^{N}
\frac{\partial^2 \psi_0^{(N)}(\x)}{\partial x_j^2}
+ \frac{\psi_0^{(N)}(\y)}{\psi_0^{(N)}(\x)}
\frac{1}{2} \sum_{j=1}^{N} \frac{\partial^2}{\partial x_j^2}
Q_N(t, \y|\x).
\label{eqn:Th1p2}
\end{equation}
Since $\psi_0^{(N)}(\x)$ is the
eigenfunction of (\ref{eqn:qToda1}) with zero eigenvalue,
\begin{equation}
\sH_N \psi_0^{(N)}(\x)
= -\frac{1}{2} \sum_{j=1}^{N}
\frac{\partial^2 \psi_0^{(N)}(\x)}{\partial x_j^2}
+\sum_{j=1}^{N-1} e^{-(x_{j+1}-x_j)} \psi_0^{(N)}(\x)=0, 
\label{eqn:Th1p3}
\end{equation}
and $Q_N(t, \y|\x)$ satisfies the equation
\begin{equation}
\frac{\partial}{\partial t} Q_N(t, \y|\x)
= \frac{1}{2} \sum_{j=1}^{N} \frac{\partial^2}{\partial x_j^2}
Q_N(t, \y|\x)
-\sum_{j=1}^{N-1} e^{-(x_{j+1}-x_j)} Q_N(t, \y|\x),
\label{eqn:Th1p4}
\end{equation}
we can see that
(\ref{eqn:Th1p2}) is equal to 
$\partial P_N(t, \y|\x)/\partial t$;
that is, $P_N(t, \y|\x)$ satisfies (\ref{eqn:OConnell1}).
Since $Q_N(0, \y|\x)=\delta(\x-\y)$, (\ref{eqn:PN3}) gives
$P_N(0,\y|\x)=\delta(\x-\y)$. 
The proof is then completed. \qed
\vskip 0.3cm

O'Connell \cite{OCo09} introduced a diffusion process
in $\R^N$ with the infinitesimal generator of the process
\begin{equation}
\sG^{\vmu}_N= \frac{1}{2} \Delta
+\nabla \log \psi_{\vmu}^{(N)}(\x) \cdot \nabla
\label{eqn:OConnell3}
\end{equation}
with $\vmu \in \R^N$.
Our theorem states that
the special case with $\vmu=0$
can be realized as a vicious BM,
which is obtained above as a system of killing BMs conditioned to survive
forever. 
Corresponding to the backward Kolmogorov
equation (\ref{eqn:OConnell1}), 
the SDEs of the process
$\Z(t)=(Z_1(t), \dots, Z_N(t)), t \geq 0$
is then given by
\begin{equation}
dZ_j(t)=dB_j(t)
+F^{(N)}_{j}(\Z(t)) dt,
\quad 1 \leq j \leq N, t \geq 0
\label{eqn:OConnell4}
\end{equation}
with
\begin{equation}
F^{(N)}_j(\x)=\frac{\partial \log \psi_0^{(N)}(\x)}
{\partial x_j},
\label{eqn:OConnell5}
\end{equation}
where $B_j(t), 1 \leq j \leq N, t \geq 0$
are independent one-dimensional BMs.

\section{Matsumoto-Yor process as a  
killing Brownian motion conditioned to survive forever}

For $N=2$, the eigenfunction (\ref{eqn:qToda7}) of
the quantum Toda lattice is
\begin{eqnarray}
&& \psi^{(2)}_{(\lambda_1, \lambda_2)}(x_1, x_2)
= \int_{\Gamma_{2}(x_1, x_2)}
e^{\sF^{(2)}_{(\lambda_1, \lambda_2)}(\T)} d \T
\nonumber\\
&& \quad
= \int_{-\infty}^{\infty} \exp \Big[
\lambda_1 T_{1,1} + \lambda_2(x_1+x_2-T_{1,1})
-\{e^{-(T_{1,1}-x_1)}+e^{-(x_2-T_{1,1})} \} 
\Big] d T_{1,1}
\nonumber\\
&& \quad
=e^{\lambda_2 (x_1+x_2)}
\int_{-\infty}^{\infty} e^{(\lambda_1-\lambda_2) T_{1,1}}
\exp \left[ -\left(
\frac{e^{x_1}}{e^{T_{1,1}}}
+\frac{e^{T_{1,1}}}{e^{x_2}} 
\right) \right]
dT_{1,1},
\label{eqn:psi2A1}
\end{eqnarray}
$(\lambda_1, \lambda_2) \in \C^2, (x_1, x_2) \in \R^2$.
Change the integral variables
$T_{1,1} \mapsto s$ by
$e^{T_{1,1}}=e^{(x_1+x_2)/2} s$.
Then
\begin{equation}
\psi^{(2)}_{(\lambda_1,\lambda_2)}(x_1,x_2)
=e^{(\lambda_1+\lambda_2)(x_1+x_2)/2}
\int_{0}^{\infty} s^{\lambda_1-\lambda_2-1}
\exp \left[ - e^{-(x_2-x_1)/2}
\left(s +\frac{1}{s} \right) \right] ds.
\label{eqn:psi2A2}
\end{equation}

Let $I_{\nu}(z)$ be the modified Bessel function
of the first kind
\begin{equation}
I_{\nu}(z)=\sum_{k=0}^{\infty} 
\frac{(z/2)^{\nu+2k}}{\Gamma(k+1) \Gamma(k+\nu+1)},
\quad |z| < \infty, \quad
|{\rm arg}z| < \pi
\label{eqn:Inu1}
\end{equation}
and $K_{\nu}(z)$ be Macdonald's function \cite{Leb65}
\begin{equation}
K_{\nu}(z)=\frac{\pi}{2}
\frac{I_{-\nu}(z)-I_{\nu}(z)}{\sin (\nu \pi)},
\quad |{\rm arg} z|< \pi, \quad \mbox{for} \quad
\nu \not= 0, \pm 1, \pm 2, \dots,
\label{eqn:Knu1}
\end{equation}
and for integers $\nu=n$,
\begin{equation}
K_{n}(z)= \lim_{\nu \to n} K_{\nu}(z),
\quad n=0, \pm 1, \pm 2, \dots,
\label{eqn:Knu2}
\end{equation}
which are both analytic functions
of $z$ for all $z$ in the complex
plane cut along the negative real axis,
and entire functions of $\nu$.
We can see that $I_{\nu}(z)$ and $K_{\nu}(z)$
are linearly independent solutions
of the differential equation
\begin{equation}
\frac{d^2 w}{dz^2}
+\frac{1}{z} \frac{dw}{dz}
-\left( 1+ \frac{\nu^2}{z^2} \right) w=0.
\label{eqn:Bessel1}
\end{equation}
For $x >0$ and $\nu \geq 0$,
$I_{\nu}(x)$ is a positive function
which increases monotonically as $x \to \infty$,
while $K_{\nu}(x)$ is a positive function
which decreases monotonically
as $x \to \infty$.
Since $K_{\nu}(z)$ has the following integral
representation
\begin{equation}
K_{\nu}(z)=\frac{1}{2} \int_0^{\infty} s^{\nu-1}
\exp \left[-\frac{z}{2}
\left( s + \frac{1}{s} \right) \right] ds,
\label{eqn:Knu3}
\end{equation}
(\ref{eqn:psi2A2}) is written as
\begin{equation}
\psi^{(2)}_{(\lambda_1, \lambda_2)}(x_1, x_2)
=2 e^{(\lambda_1+\lambda_2)(x_1+x_2)/2}
K_{\lambda_1-\lambda_2}(2 e^{-(x_2-x_1)/2} ).
\label{eqn:psi2A3}
\end{equation}

The infinitesimal generator of the process
(\ref{eqn:OConnell3}) is then given for $N=2$ as
\begin{eqnarray}
\sG^{(\mu_1, \mu_2)}_2
&=& \frac{1}{2} \Delta
+\sum_{j=1}^{2} \frac{\partial}{\partial x_j} 
\log \psi^{(2)}_{\lambda_1, \lambda_2}(x_1, x_2)
\frac{\partial}{\partial x_j}
\nonumber\\
&=& \frac{1}{2} \left( \frac{\partial^2}{\partial x_1^2}
+\frac{\partial^2}{\partial x_2^2} \right)
+\frac{1}{2}(\mu_1+\mu_2)
\left( \frac{\partial}{\partial x_1}
+\frac{\partial}{\partial x_2} \right)
\nonumber\\
&& + \frac{K'_{\mu_1-\mu_2}(2e^{-(x_2-x_1)/2})}
{K_{\mu_1-\mu_2}(2e^{-(x_2-x_1)/2})}
e^{-(x_2-x_1)/2}
\left( \frac{\partial}{\partial x_1}
-\frac{\partial}{\partial x_2} \right),
\label{eqn:GN2_1}
\end{eqnarray}
where $K'_{\nu}(z) \equiv dK_{\nu}(z)/dz$.
If we change the variables
$(x_1, x_2) \mapsto (\xi, \eta)$
by $\xi=(x_1+x_2)/2$, $\eta=-(x_1-x_2)/2-\log 2$,
(\ref{eqn:GN2_1}) is decomposed into two parts,
$\sG^{(\mu_1, \mu_2)}_2=(\sG^{\mu_1+\mu_2}_0
+\sG^{\mu_1-\mu_2}_{\rm MY})/2$, where
\begin{eqnarray}
\label{eqn:G0}
\sG_0^{\nu}
&=& \frac{1}{2} \frac{d^2}{d \xi^2}+\nu \frac{d}{d \xi},
\\
\label{eqn:GMY}
\sG_{\rm MY}^{\mu}
&=& \frac{1}{2} \frac{d^2}{d \eta^2}
+\frac{d}{d \eta} \{ \log K_{\mu}(e^{-\eta}) \}
\frac{d}{d \eta}
= \frac{1}{2} \frac{d^2}{d \eta^2}
- \frac{K'_{\mu}(e^{-\eta})}{K_{\mu}(e^{-\eta})} e^{-\eta}
\frac{d}{d \eta}.
\end{eqnarray}
The former is the infinitesimal generator
of the one-dimensional BM with a constant drift
$\nu=\mu_1+\mu_2$
and the latter is that of the diffusion process
studied by Matsumoto and Yor 
with parameter $\mu=\mu_1-\mu_2$ \cite{MY00,MY05}.
It implies that, in the $N=2$ case
of the O'Connell process with parameter
$\vmu=(\mu_1, \mu_2) \in \R^2$,
the center of mass $\xi$ behaves as
(a time change $t \mapsto 2t$ of)
a BM with a drift $\mu_1+\mu_2$
and the relative coordinate $\eta$
behaves as (a time change $t \mapsto 2t$ of)
the Matsumoto-Yor process with parameter
$\mu_1-\mu_2$.

For $\mu \in \R$, let
\begin{equation}
\sL^{\mu}
= \frac{1}{2} \frac{d^2}{dx^2}-V(x)-\mu \frac{d}{dx}
\label{eqn:MYa1}
\end{equation}
with
\begin{equation}
V(x)=\frac{1}{2} e^{-2x}.
\label{eqn:MYa2}
\end{equation}
It is the infinitesimal generator of the
one-dimensional drifted BM with
a killing term $-V(x)$.
The transition probability density is given by
\begin{eqnarray}
Q^{\mu}(t, y|x)
&=& \bE \left[ \1(B^{\mu,y}(t)=x)
\exp \left\{-\frac{1}{2} \int_0^{t} e^{-2 B^{\mu,y}(s)} ds \right\} \right]
\nonumber\\
&=& e^{-\mu^2 t/2+\mu(x-y)}
\frac{i}{\pi^2} \int_{-i \infty}^{i \infty}
e^{\lambda^2 t/2} K_{\lambda}(e^{-x})
K_{\lambda}(e^{-y}) 
\lambda \sin(\pi \lambda) d \lambda,
\label{eqn:MYa3}
\end{eqnarray}
where $B^{\mu,y}(t)=y+B(t)+\mu t$ with
the one-dimensional standard BM, $B(t)$,
starting from 0; $B(0)=0$.

The survival probability up to time $T >0$
of this one-dimensional killing BM
with drift $\mu$ is given by
\begin{equation}
\sN^{\mu}(T, x)
=\int_{-\infty}^{\infty} Q^{\mu}(T, y|x) dy
\label{eqn:MYa4}
\end{equation}
for the initial position $x \in \R$.
We can prove that, if $\mu>0$, it has the
long-term asymptotics, 
\begin{eqnarray}
\lim_{T \to \infty} 
\sqrt{\frac{\pi}{2}} T^{3/2}
e^{\mu^2 T/2}
\sN^{\mu}(T,x)
&=& e^{\mu x} K_{0}(e^{-x})
\int_{-\infty}^{\infty} K_0(e^{-y}) e^{-\mu y} dy
\nonumber\\
&=& 2^{\mu-2} (\Gamma(\mu/2))^2 e^{\mu x} K_0(e^{-y}).
\label{eqn:MYa5}
\end{eqnarray}
The proof of (\ref{eqn:MYa5}) is given in
Appendix \ref{chap:appendix_MY}.
Then the transition probability density for this
{\it one-dimensional killing BM conditioned to survive forever}
is given in the limit $\mu \to 0, \mu >0$ as
\begin{eqnarray}
P(t, y|x)
&=& \lim_{\mu \to 0, \mu >0} \lim_{T \to \infty}
\frac{\sN^{\mu}(T-t,y)}{\sN^{\mu}(T,x)}
Q^{\mu}(t, y|x)
\nonumber\\
&=& \frac{K_0(e^{-y})}{K_0(e^{-x})}
\frac{i}{\pi^2}
\int_{-i \infty}^{i \infty} e^{\lambda^2 t/2}
K_{\lambda}(e^{-x}) K_{\lambda}(e^{-y})
\lambda \sin(\pi \lambda) d \lambda,
\label{eqn:MYa6}
\end{eqnarray}
$x,y \in \R, t > 0$.
It is easy to see that (\ref{eqn:MYa6})
satisfies the diffusion equation
\begin{equation}
\frac{\partial}{\partial t} u(t, x)
=\sG^0_{\rm MY} u(t, x),
\quad t \geq 0,
\label{eqn:MYa7}
\end{equation}
under the initial condition
$u(0,x)=\delta(x-y)$.

Matsumoto and Yor showed that the stochastic
process
\begin{equation}
Z^{\mu}_{\rm MY}(t)
= \log \left\{ \int_0^t e^{2 B^{\mu}(s)} ds \right\}
-B^{\mu}(t), \quad t \geq 0
\label{eqn:MYb1}
\end{equation}
with $B^{\mu}(t) \equiv B^{\mu,0}(t)=B(t)+\mu t$
is a diffusion process, whose infinitesimal generator
is given by (\ref{eqn:GMY}) for any $\mu \in \R$.
Here we have shown that, 
when $\mu=0$, the Matsumoto-Yor process (\ref{eqn:MYb1})
can be constructed as a one-dimensional
killing BM conditioned to survive forever.

See Appendix B for more detail on the relation
between the Matsumoto-Yor process and the
$N=2$ case of the O'Connell process.

\section{Special Initial Conditions}

In this section, first we consider the one-dimensional
diffusion process with the infinitesimal generator (\ref{eqn:MYa1})
with (\ref{eqn:MYa2}).
Let $0 < T < \infty$.
Then the transition probability density of the process
conditioned to survive up to time $T$ is given by
\begin{eqnarray}
P_T^{\mu}(s,x; t,y) &=&
\frac{\sN^{\mu}(T-t,y)}{\sN^{\mu}(T-s,x)}
Q^{\mu}(t-s,y|x)
\nonumber\\
&=& \sN^{\mu}(T-t,y)
\frac{Q^{\mu}(t-s,y|x)}
{\int_{-\infty}^{\infty} Q^{\mu}(T-s,z|x) dz},
\label{eqn:PTB1}
\end{eqnarray}
$0 \leq s \leq t \leq T, x,y \in \R$, where
$Q^{\mu}$ and $\sN^{\mu}$ are given by
(\ref{eqn:MYa3}) and (\ref{eqn:MYa4}), respectively.
The asymptotics of $K_{\lambda}(e^{-x})$ in $x \to - \infty$
is independent of $\lambda$ \cite{Leb65} 
\begin{equation}
K_{\lambda}(e^{-x})
\simeq \sqrt{\frac{\pi}{2 e^{-x}}} 
\exp(-e^{-x})
\quad \mbox{as $x \to - \infty$}.
\label{eqn:asymK}
\end{equation}

Then if we define
\begin{equation}
P_T^{\mu}(t, y|-\infty) \equiv
\lim_{x \to -\infty} P_T^{\mu}(0, x; t,y),
\label{eqn:PTB2}
\end{equation}
it is given by
\begin{eqnarray}
P_T^{\mu}(t,y|-\infty)
&=& \frac{\sqrt{2 \pi}}{2^{\mu-2}(\Gamma(\mu/2))^2} T^{3/2}
e^{\mu^2(T-t)/2 -\pi^2/T}
\theta_{e^{-y}}(t) e^{-\mu y}
\sN(T-t,y),
\label{eqn:PTB3}
\end{eqnarray}
where $\theta_r(t)$ is given by (\ref{eqn:Yor1})
and (\ref{eqn:Yor2}) in Appendix A \cite{Yor01}.
Since $\sN^{\mu}(0,y)=1$ by definition,
we obtain the distribution at time $t=T$,
\begin{equation}
P_T^{\mu}(T,y|-\infty)
=c^{\mu}(T) \theta_{e^{-y}}(T) e^{-\mu y},
\quad y \in \R, \quad T > 0
\label{eqn:PTB4}
\end{equation}
with $c^{\mu}(T)=\sqrt{2\pi} 2^{2-\mu}(\Gamma(\mu/2))^{-2} T^{3/2} e^{-\pi^2/T}$.
On the other hand, by (\ref{eqn:MYa5}), if we take
the temporally homogeneous limit $T \to \infty$
in (\ref{eqn:PTB3}), then we obtain the 
distribution
\begin{eqnarray}
P^{\mu}(t,y|-\infty)
&\equiv& \lim_{T \to \infty} P_T^{\mu}(t,y|-\infty)
\nonumber\\
&=& 2 e^{-\mu^2 t/2}
\theta_{e^{-y}}(t) K_0(e^{-y}),
\quad y \in \R, \quad t > 0.
\label{eqn:PTB5}
\end{eqnarray}

Next we consider the $N$-particle system of the
killing BMs with drift $\vmu \in \W_N$ conditioned
to survive up to time $T, 0 < T < \infty$.
The transition probability density is given by
(\ref{eqn:PN1}).
Corresponding to (\ref{eqn:asymK}),
the asymptotics of $\psi^{(N)}_{\vlambda}(\x)$ in
$x_j \to -\infty, 1 \leq \forall j \leq N$
is independent of $\vlambda$ (see Remark 8.1 in \cite{OCo09}).
Then we obtain
\begin{eqnarray}
P^{\vmu}_{N,T}(t,\y|-\infty)
&\equiv& \lim_{x_j \to -\infty, 1 \leq j \leq N}
P^{\vmu}_{N,T}(0, \x; t, \y)
\nonumber\\
&=& \frac{e^{|\vmu|^2(T-t)/2}}
{J^{\vmu}(N,T)} 
\Theta_N(t,\y) \exp (-\vmu \cdot \y)
\sN_N^{\vmu}(T-t,\y),
\label{eqn:PTC1}
\end{eqnarray}
where
\begin{equation}
J^{\vmu}(N,T)=\int_{(i \R)^N} 
e^{T \sum_{j=1}^N \lambda_j^2 /2} s_N(\vlambda)
I^{\vmu}_{-\vlambda}(N) d \vlambda
\label{eqn:PTC2}
\end{equation}
with
\begin{equation}
I^{\vmu}_{-\vlambda}(N)
=\int_{\R^N} \psi^{(N)}_{-\vlambda}(\z)
\exp(-\vmu \cdot \z) d\z,
\quad \vlambda \in (i \R)^N,
\quad \vmu \in \W_N, 
\label{eqn:PTC3}
\end{equation}
and \cite{OCo09}
\begin{equation}
\Theta_{N}(t, \y)= \int_{(i \R)^N} 
e^{t \sum_{j=1}^{N} \lambda_j^2/2} \psi_{-\vlambda}^{(N)}(\y)
s_N(\vlambda) d \vlambda.
\label{eqn:PTC4}
\end{equation}
At time $t=T$, (\ref{eqn:PTC1}) gives the distribution
\begin{equation}
P^{\vmu}_{N,T}(T, \y|-\infty)
=\frac{1}{J^{\vmu}(N,T)}
\Theta_N(T, \y) \exp (-\vmu \cdot \y),
\quad \y \in \R^N, \quad T >0.
\label{eqn:PTC5}
\end{equation}
On the other hand, if we take the limit
$T \to \infty$, (\ref{eqn:PTC1}) gives
the distribution
\begin{eqnarray}
P^{\vmu}_N(t, \y|-\infty)
&\equiv& \lim_{T \to \infty}
P^{\vmu}_{N,T}(t, \y|-\infty)
\nonumber\\
&=& e^{-|\vmu|^2 t/2}
\Theta_N(t,\y) \psi_0^{(N)}(\y),
\quad \y \in \R^N, \quad t > 0.
\label{eqn:PTC6}
\end{eqnarray}
The three-dimensional
Bessel process is defined as the radial part of the three-dimensional
BM and abbreviated to BES(3).
The above results will be compared with 
the Imhof relation between BES(3) and 
the process called a meander and its multivariate
generalizations discussed in \cite{KT02,KT04}.

\section{Summary and Concluding Remarks}

The vicious BM is obtained as a diffusion
scaling limit of Fisher's vicious walk model 
\cite{Fis84,KT02,CK03}.
It is an $N$-particle system of BMs 
in one dimension,
whose positions are arranged in the order
$x_1<x_2< \cdots < x_N$, such that
if and only if two neighboring Brownian particles
collide with each other then they are pair
annihilated, while they can enjoy free Brownian motions
if they are all located separately from each other.
In the present paper we have considered a system of
$N$ Brownian particles with the killing term
\begin{equation}
-V_N(\x)=-\sum_{j=1}^{N-1} e^{-(x_{j+1}-x_j)}.
\label{eqn:VNB1}
\end{equation}
That is, the interactions between neighboring
Brownian particles are long-ranged and 
the risk to be pair annihilated exists always, 
which is expressed by a rapid decreasing function 
(\ref{eqn:VNB1}) of the distance of
the two particles $x_{j+1}-x_j$.
We regard this system of mutually
killing BMs as a generalized version of
vicious BM, since the original vicious BM
can be identified with the system of BMs 
with the killing term obtained by 
$-\lim_{\varepsilon \to 0, \varepsilon >0} 
V_N(\x/\varepsilon)$.

Though the original vicious BM has only contact
interactions, if we consider the system 
conditioned never to collide with each other, 
then we obtain a system of BMs
with long-ranged interactions;
the SDE is given by Eq.(\ref{eqn:Dyson1}),
in which between any pair of particles
there acts a repulsive force proportional
to the inverse of distance of the pair \cite{KT07}.
This $N$-particle process is equivalent to
the eigenvalue process of an $N \times N$
Hermitian-matrix-valued BM introduced by
Dyson in order to dynamically simulate
the eigenvalue statistics of the Gaussian unitary
ensemble (GUE) of random matrices
(the Dyson model) \cite{Dys62,Meh04,For10}.

As discussed in \cite{KT_Sugaku,Katori_Oka},
the equivalence between the eigenvalue process
of Dyson and the noncolliding BM
(the vicious BM conditioned never to collide)
is the $N$-variate extension of the 
equivalence between BES(3) and the one-dimensional BM
conditioned to stay positive.
In this sense, the Dyson model can be
regarded as a many-particle generalization
of BES(3).

Apart from the equivalence between the BES(3)
and the conditional BM to stay positive,
the following equivalence is established.
Let $M(t)=\max_{0 \leq s \leq t} B(s), t \geq 0$,
and define a process
$Y(t)=2M(t)-B(t), t \geq 0$.
Then $Y(t)$ is equivalent to BES(3),
which is known as Pitman's `$2M-X$' theorem
\cite{Pit75} (see also \cite{RP81,MY00,MY05}).
As a multivariate extension of
Pitman's `$2M-X$' theorem, another construction
of the Dyson model (the noncolliding BM)
has been reported \cite{BJ02,OY02,War07,FN08,BFPSW09}.

Matsumoto and Yor studied the stochastic process
$Z^{\mu}_{\rm MY}(t), t \geq 0$ given by
(\ref{eqn:MYb1}). We can see that
\begin{eqnarray}
&&\lim_{\varepsilon \to 0, \varepsilon >0} 
\varepsilon Z_{\rm MY}^{\mu}(t/\varepsilon^2)
\nonumber\\
&& = \lim_{\varepsilon \to 0, \varepsilon > 0}
\left[ \varepsilon \log \left\{
\int_0^{t} e^{2 B^{\mu}(s)/\varepsilon} ds \right\}
-B^{\mu}(t)- \varepsilon \log \varepsilon^2 \right]
\nonumber\\
&& = 2 \max_{0 \leq s \leq t} B^{\mu}(s)-B^{\mu}(t),
\quad t \geq 0.
\label{eqn:MYc1}
\end{eqnarray}
Then, when $\mu=0$,
this $\varepsilon \to 0$ limit is equivalent to
$Y(t)$ and thus with the BES(3).
In this sense, the Matsumoto-Yor process
is a generalization of the BES(3) \cite{MY00,MY05}.

O'Connell \cite{OCo09} introduced
an $N$-particle process
$\Z^{\vmu}(t)=(Z_1^{\vmu}(t), \dots, Z_N^{\vmu}(t)),
t \geq 0$,$\vmu \in \R^N$ as a multi-dimensional generalization
of the Matsumoto-Yor process.
Corresponding to the fact that
the Matsumoto-Yor process is a generalization
of the BES(3), 
the O'Connell process is a generalization
of the Dyson model.
Actually, he showed that
$\lim_{\varepsilon \to 0, \varepsilon>0} 
\varepsilon \Z^{0}(t/\varepsilon^2), t \geq 0$
is equivalent to the Dyson model
in the sense of an extension of
Pitman's `$2M-X$' theorem \cite{OCo09}.
We pointed out that 
the BES(3), 
the original vicious BM, and
the Dyson model (the noncolliding BM)
can be regarded as 
{\it ultradiscretizations}
\cite{TTMS96}
of the Matsumoto-Yor process,
the BMs with the 
killing term in the same form as
the quantum Toda lattice potential, 
and the O'Connell process, respectively.

In the present paper, we discussed 
another construction of the O'Connell process
apart from the extension of 
Pitman's `$2M-X$' theorem.
In the special case with $\vmu=0$, we have
shown here that his process is given as
a generalized version of vicious BM 
conditioned to survive forever.
In order to demonstrate that
the relation between the present generalized vicious BM
and the O'Connell process
is a multivariate generalization of 
the relation between a killing BM and 
the Matsumoto-Yor process, 
we showed in Sec.V that
the Matsumoto-Yor process with $\mu=0$
is obtained as a killing BM conditioned to survive forever.

We want to emphasize that
the present analysis is indeed based on
the idea of O'Connell to discuss 
interacting diffusive particle systems
using the exact solutions of
the quantum Toda lattices \cite{OCo09}.

In Sec.I in the present paper, we listed up
three fundamental properties of the 
noncolliding BM. They are all inherited by
the O'Connell process in the extended form.
(i) The Karlin-McGregor determinantal expression (\ref{eqn:KM1})
of $q_N(t,\y|\x)$, which is expanded by
the Schur functions (\ref{eqn:qNexp1}),
is generalized by the integral formula 
(\ref{eqn:QNexp1}).
(ii) The harmonic transform 
\cite{Doo84} from $q_N$ to $p_N$ (\ref{eqn:h-trans1})
by the harmonic function $h_N$ given by
the product of differences of variables
(the Vandermonde determinant) (\ref{eqn:hN1})
is now given by the formula (\ref{eqn:PN3})
from $Q_N$ to $P_N$. There $h_N$ is
replaced by an eigenfunction $\psi^{(N)}_0$
of the infinitesimal generator $\sL_N$
(the Hamiltonian $\sH_N$ of the quantum Toda lattice).
(See \cite{Mae06,Tak10,TT11} for harmonic transforms of 
one dimensional generalized diffusion processes.)
(iii) Theorem 2 in Sec.IV gives the extended 
version of the Kolmogorov equation of (\ref{eqn:Kolm1}).

There are a lot of future problems.
In noncolliding diffusion processes,
if we study the situations starting from
``the all zero state'' and observe
particle distributions at an arbitrary
time $0<t<\infty$ in temporally homogeneous processes, 
and at the ending time $t=T$ in temporally
inhomogeneous processes
defined only in an finite time-interval 
$[0, T]$, we have obtained the eigenvalue
distributions of random matrices
in a variety of ensembles
studied in random matrix theory \cite{KT04}.
In Sec.VI, we demonstrated that
``the all $-\infty$ state'' and 
temporally inhomogeneous versions of
processes will play important roles
in the O'Connell process.
The noncolliding diffusion process is
determinantal, in the sense that
for any finite initial configuration
all multitime correlation functions are
given by determinants associate with
an integral kernel called the
correlation kernel \cite{KT10,KT11}.
It will be a challenging problem
to clarify how matrix-structures
({\it i.e.} symmetries of systems) \cite{Meh04,For10}
and solvability are inherited
by the family of O'Connell processes.

\begin{acknowledgments}
The present author would like to thank
H. Tanemura, T. Sasamoto, T. Imamura
for useful discussions on the present work.
A part of the present work was done
during the participation of the present author 
in \'Ecole de Physique des Houches on
`` Vicious Walkers and Random Matrices"
(May 16-27, 2011).
The author thanks G. Schehr, C. Donati-Martin, and
S. P\'ech\'e 
for well-organization of the school
and R. Chhaibi for useful discussion 
on the Matsumoto-Yor process.
This work is supported in part by
the Grant-in-Aid for Scientific Research (C)
(No.21540397) of Japan Society for
the Promotion of Science.
\end{acknowledgments}

\appendix
\section{Proof of (\ref{eqn:MYa5})}
\label{chap:appendix_MY}

Let $J_0(z)$ be the Bessel function of the first kind
of order 0,
\begin{equation}
J_0(z)=\sum_{k=0}^{\infty} 
\frac{(-1)^k (z/2)^{2k}}{(k!)^2}, \quad
|z| < \infty.
\label{eqn:J0}
\end{equation}
Since the equality
\begin{equation}
K_{\lambda}(x) K_{\lambda}(y)
=\frac{\pi}{2 \sin(\pi \lambda)}
\int_{\log(y/x)}^{\infty}
J_0(\sqrt{2 xy \coth u- x^2-y^2})
\sinh (u \lambda) du
\label{eqn:KK1}
\end{equation}
holds for $x>0, y>0, |{\rm Re} \lambda| < 1/4$
\cite{Leb65},
(\ref{eqn:MYa3}) is written as
\begin{eqnarray}
&& Q^{\mu}(t, y|x)
= e^{-\mu^2 t/2 + \mu(x-y)}
\frac{i}{2 \pi^2}
\int_{x-y}^{\infty} du \,
J_0(\sqrt{2 e^{-(x+y)} \cosh u - e^{-2x}-e^{-2y}})
\nonumber\\
&& \qquad \qquad \times
\int_{-i \infty}^{i \infty} d \lambda \,
e^{\lambda^2 t/2} \lambda \sinh(u \lambda)
\nonumber\\
&& \quad =
\frac{e^{-\mu^2 t/2+\mu(x-y)}}{\sqrt{2 \pi} t^{3/2}}
\int_{x-y}^{\infty} u
J_0(\sqrt{2 e^{-(x+y)} \cosh u - e^{-2x}-e^{-2y}})
e^{-u^2/2t} du,
\label{eqn:KK2}
\end{eqnarray}
where we have performed the integral of
$\lambda$ over $i \R$.
This gives
\begin{eqnarray}
&& \lim_{t \to \infty}
\sqrt{\frac{\pi}{2}} t^{3/2} e^{\mu^2 t/2}
Q^{\mu}(t, y|x)
\nonumber\\
&& \quad = e^{\mu(x-y)}
\frac{1}{2} \int_{x-y}^{\infty} u 
J_0(\sqrt{2 e^{-(x+y)} \cosh u - e^{-2x}-e^{-2y}}) du.
\label{eqn:KK3}
\end{eqnarray}
We find that the $\lambda \to 0$ limit of 
(\ref{eqn:KK1}) gives the equality
\begin{equation}
K_0(x) K_0(y)
=\frac{1}{2} \int_{\log(y/x)}^{\infty} u
J_0(\sqrt{2 e^{-(x+y)} \cosh u - e^{-2x}-e^{-2y}}) du,
\label{eqn:KK4}
\end{equation}
and then (\ref{eqn:KK3}) gives
\begin{equation}
\lim_{t \to \infty}
\sqrt{\frac{\pi}{2}} t^{3/2} e^{\mu^2 t/2}
Q^{\mu}(t, y|x)
=e^{\mu(x-y)} K_{0}(e^{-x}) K_{0}(e^{-y}).
\label{eqn:KK5}
\end{equation}
The asymptotics of $K_0(e^{-y})$
is known as \cite{Leb65}
\begin{equation}
K_0(e^{-y}) \simeq \left\{
\begin{array}{ll}
\log (2/e^{-y}) \sim y &
\quad \mbox{as $y \to \infty$} \cr
\sqrt{\pi/(2 e^{-y})} \exp(-e^{-y})
\to 0 &
\quad \mbox{as $y \to -\infty$}.
\end{array} \right.
\label{eqn:KK6}
\end{equation}
Then if $\mu > 0$,
$\int_{-\infty}^{\infty} K_0(e^{-y}) e^{-\mu y} dy < \infty$.
Actually, for $\mu >0$, this integral is the
Mellin transformation of $K_0(z)$
and we obtain
\begin{equation}
\int_{-\infty}^{\infty} K_0(e^{-y}) e^{-\mu y} dy
=\int_0^{\infty} K_0(z) z^{\mu-1} dz
= 2^{\mu-2}(\Gamma(\mu/2))^2.
\label{eqn:KK7}
\end{equation}
Therefore (\ref{eqn:MYa5}) is valid. \qed
\vskip 0.5cm

We note that, for the function
\begin{equation}
\theta_r(t)=\frac{i}{2 \pi^2}
\int_{-i \infty}^{i \infty} e^{\lambda^2 t/2}
K_{\lambda}(r) \lambda \sin(\pi \lambda) d \lambda,
\quad r > 0,
\label{eqn:Yor1}
\end{equation}
Yor gave the following expression
(see Eq.(6.b'') on page 43 of \cite{Yor01}),
\begin{equation}
\theta_r(t) = \frac{r}{(2 \pi^3 t)^{1/2}} e^{\pi^2/2t}
\int_0^{\infty} e^{-\eta^2/2t} e^{-r \cosh \eta}
(\sinh \eta) \sin \left( \frac{\pi \eta}{t} \right) d\eta,
\quad r > 0.
\label{eqn:Yor2}
\end{equation}
Using this expression, Matsumoto and Yor
reported the asymptotics
(see Eq.(2.11) in \cite{MY05}),
\begin{equation}
\lim_{t \to \infty}
\sqrt{ 2 \pi t^3} \theta_r(t)
=K_{0}(r), \quad r > 0.
\label{eqn:Yor3}
\end{equation}
Since the equality
\begin{equation}
Q^{\mu}(t,y|x)
=e^{-\mu^2 t/2 +\mu(x-y)}
\int_0^{\infty} \exp \left\{
-\frac{s}{2}-\frac{1}{2s}
(e^{-2x}+e^{-2y}) \right\}
\theta_{e^{-(x+y)}/s}(t) \frac{ds}{s}
\label{eqn:Yor4}
\end{equation}
is established, the limit (\ref{eqn:KK5})
can be concluded also from (\ref{eqn:Yor3}).

\section{$N=2$ case of the O'Connell process}
\label{chap:N=2case}

By the equations (\ref{eqn:Skl1}), (\ref{eqn:tpd1}) and 
(\ref{eqn:psi2A3}), we obtain
\begin{eqnarray}
Q_2(t, \y|\x)
&=& \frac{1}{2 \pi^3}
\int_{-i \infty}^{i \infty} d \lambda_1 
\int_{-i \infty}^{i \infty} d \lambda_2 \,
e^{(\lambda_1^2+\lambda_2^2)t/2}
e^{(\lambda_1+\lambda_2)\{(x_1+x_2)-(y_1+y_2)\}/2}
\nonumber\\
&& \times
K_{\lambda_1-\lambda_2}(2 e^{-(x_2-x_1)/2})
K_{\lambda_1-\lambda_2}(2 e^{-(y_2-y_1)/2})
(\lambda_1-\lambda_2) \sin\{\pi(\lambda_1-\lambda_2)\},
\label{eqn:Q2B1}
\end{eqnarray}
$\x, \y \in \R^2, t \geq 0$.
If we change the integral variables
$(\lambda_1, \lambda_2) \mapsto
(\lambda, \nu)$ by 
$\lambda=\lambda_1-\lambda_2, \nu=\lambda_1+\lambda_2$,
we can calculate the integral with respect to $\nu$.
The result is expressed by using the 
transition probability density $Q^{\mu}$ of the
Matsumoto-Yor process (\ref{eqn:MYa3}) with $\mu=0$ as
\begin{equation}
Q_2(t, \y|\x)
=p(2t,y_1+y_2|x_1+x_2)
Q^0 \left( \frac{t}{2}, \frac{y_2-y_1}{2}-\log 2
\Big| \frac{x_2-x_1}{2}-\log 2 \right),
\label{eqn:Q2B2}
\end{equation}
where $p(t,y|x)=e^{-(y-x)^2/2t}/\sqrt{2 \pi t}$.
Therefore, from the long-term asymptotics
(\ref{eqn:KK5}) of $Q^{\mu}$, we can obtain the
long-term asymptotics of $Q_2$ as
\begin{eqnarray}
Q_2(t,\y|\x)
&\simeq& \frac{2}{\pi t^2} K_0(2 e^{-(x_2-x_1)/2})
K_0(2 e^{-(x_2-x_1)/2})
\nonumber\\
&=& \frac{t^{-2}}{2\pi} \psi^{(2)}_0(\x)
\psi_0^{(2)}(\y)
\quad \mbox{as $t \to \infty$},
\label{eqn:Q2B3}
\end{eqnarray}
which coincides with the $N=2$ case
of (\ref{eqn:asymB2}) in Lemma 1.



\end{document}